\documentclass[12pt,tightenlines,eqsecnum,floats,aps,amsmath,amssymb,nofootinbib,superscriptaddress,showpacs,prd]{revtex4}

\usepackage{dcolumn}
\usepackage{bm}
\usepackage[spanish,english]{babel}
\usepackage{amsfonts}
\usepackage{amssymb}
\usepackage{graphicx}
\usepackage[latin1]{inputenc}

\newcommand{\Hcal}{\mathcal{H}}

\begin{document}

\title{ {\bf Isotropic and Anisotropic Bouncing Cosmologies in Palatini Gravity}}

\author{Carlos Barragán}
\affiliation{ \footnotesize Departamento de Física Teórica, Universidad Autónoma de Madrid,  28049 Madrid, Spain}

\author{Gonzalo J. Olmo}\email{olmo@iem.cfmac.csic.es}
\affiliation{ {\footnotesize Instituto de Estructura de la Materia, CSIC, Serrano 121, 28006 Madrid, Spain} \\ and \\
{\footnotesize Physics Department, University of Wisconsin-Milwaukee, P.O.Box 413, Milwaukee, WI 53201 USA}}

\date{May 20$^{th}$, 2010}

\begin{abstract}
We study isotropic and anisotropic (Bianchi I) cosmologies in Palatini $f(R)$ and $f(R,R_{\mu\nu}R^{\mu\nu})$ theories of gravity and consider the existence of non-singular bouncing solutions in the early universe. We find that all $f(R)$ models with isotropic bouncing solutions develop shear singularities in the anisotropic case. On the contrary, the simple quadratic model $R+a R^2/R_P+R_{\mu\nu}R^{\mu\nu}/R_P$ exhibits regular bouncing solutions in both isotropic and anisotropic cases for a wide range of equations of state, including dust (for $a<0$) and radiation (for arbitrary $a$). It thus represents a purely gravitational solution to the big bang singularity and anisotropy problems of general relativity without the need for exotic ($w>1$) sources of matter/energy. 
\end{abstract}

\pacs{04.50.Kd, 98.80.-k, 98.80.Qc}
\keywords{Modified Gravity, Palatini Formalism, Non-singular Cosmologies }

\maketitle
\newpage
\tableofcontents

\section{Introduction}

Ever since its publication, Einstein's theory of general relativity (GR) has fascinated theoretical physicists. Not only it is in excellent quantitative agreement with all observations \cite{Will05} (if a cosmological constant is included),  but it also allows us to determine at which stage it should not be trusted. The existence of cosmological (big bang) and black hole singularities are clear symptoms that the theory is not complete. To overcome this drawback, it is common to argue that in such extreme scenarios quantum gravitational effects should play an important role and would avoid the breakdown of predictability, i.e., the disappearance of physical laws. The details of how this should actually happen is another mystery and probably different quantum theories of gravity would lead to different mechanisms for removing the singularities. \\
\indent If we accept that the idea of gravitation as a geometric phenomenon still persists at the quantum level \footnote{Note that string theory predicts the existence of fields with couplings that violate the equivalence principle \cite{Will05}. For this reason, the idea of gravitation as a purely geometric phenomenon is explicitly broken in that context.} 
, with perhaps quantized areas and volumes in the fashion of loop quantum gravity \cite{LQG}, it seems reasonable to expect that the quantum corrected gravitational dynamics could be described in terms of some effective action incorporating a number of regulating parameters (while keeping the matter sector untouched). In the absence of fully understood quantum theories of gravity and of their corresponding effective actions, it would be desirable to have a working model which, as an intermediate step between  classical GR and the {\it final } quantum theory of gravity, could capture, at least qualitatively, some aspects of the sought non-singular theory of gravity.  Stated differently, can we find a {\it regulated } gravitational theory (free from singularities) and as successful as GR at low energies? Obviously, such a theory would be very welcome from a phenomenological point of view and could provide new insights on fundamental properties of the geometry at very high energies. 

In this work we elaborate in this direction and propose a family of modified Lagrangians which departs from GR by quadratic curvature corrections 
\begin{equation}\label{eq:f(R,Q)}
f(R,R_{\mu\nu}R^{\mu\nu})=R+a\frac{R^2}{R_P}+b\frac{R_{\mu\nu}R^{\mu\nu}}{R_P} \ ,
\end{equation}
where $R_P\sim l_P^{-2}$ is the Planck curvature, and show that for a wide range of parameters $a$ and $b$ they lead to non-singular cosmologies both in isotropic and anisotropic Bianchi I universes for all reasonable sources of matter and energy. In particular, we find that radiation dominated universes are always non-singular. The novelty of our approach, obviously, is not the particular Lagrangian considered, which is well known and naturally arises in perturbative approaches to quantum gravity. The new ingredient that makes our model so successful in removing cosmological singularities is the fact that we follow a first order (Palatini) formulation of the theory, in which metric and connection are assumed to be independent fields. In this approach, the metric satisfies second-order partial differential equations, like in GR, and the independent connection does not introduce any additional dynamical degrees of freedom (like in the Palatini version of GR \cite{MTW-Wald}). In fact, the connection  can be expressed in terms of the metric, its first derivatives, and functions of the matter fields and their first derivatives. As a result, the theory is identical to GR in vacuum but exhibits different dynamics when matter and/or radiation are present. For the model (\ref{eq:f(R,Q)}), this means that the dynamics is identical to that of GR at low curvatures but departures arise at high energies/curvatures. Since the equations are of second-order, there can not be more solutions in this theory than there are in GR. Therefore, the modified solutions that we find represent deformations (at the Planck scale) of the solutions corresponding to GR. Such deformations, as we will see, are able to avoid the big bang singularity by means of a bounce from an initially contracting phase to the current expanding universe. 

Our approach is motivated by previous studies on non-singular bouncing cosmologies initiated in \cite{Olmo-Singh09} and continued in \cite{BOSA09} and \cite{OSAT09}. In \cite{Olmo-Singh09} it was shown that the effective dynamics of loop quantum cosmology \cite{lqc}, which describes an isotropic bouncing universe, can be exactly derived from an $f(R)$ action with high curvature corrections in Palatini formalism. Different attempts to find effective actions for those equations followed that work but either failed \cite{Baghram:2009we} or are limited to the low-energy, perturbative regime \cite{Sot09}. The existence and characterization of bouncing cosmologies in the $f(R)$ Palatini framework was studied in \cite{BOSA09}, and the evolution of cosmological perturbations has been recently considered in \cite{Koivisto10}. The field equations of extended Palatini theories $f(R,Q)$, in which the gravity Lagrangian is also a function of the squared Ricci tensor $Q=R_{\mu\nu}R^{\mu\nu}$, were investigated in \cite{OSAT09}, where it was found that in models of the form $f(R,Q)=\tilde f(R)+Q/R_P$, the scalar $Q$ is generically bounded from above irrespective of the symmetries of the theory. Unlike in the more conventional metric formalism, Palatini Lagrangians of the form $f(R,Q)$ lead to second-order equations for the metric and, therefore, are free from ghosts and other instabilities for arbitrary values of the parameters $a$ and $b$.

\indent The successful results of \cite{Olmo-Singh09} and \cite{BOSA09} motivate and force us to explore scenarios with less symmetry to see if Palatini theories are generically an appropriate framework for the construction of non-singular theories. We will see that $f(R)$ models which lead to bouncing cosmologies in the isotropic case also lead to anisotropic Bianchi I universes with expansion and energy density bounded from above. As we show here, however, such models have an unavoidable shear divergence, which occurs when the condition $\partial_R f(R)=0$ is met. This important result implies that Palatini $f(R)$ theories do not have the necessary ingredients to allow for a fully successful {\it regulated} theory in the sense defined above. Such limitation, however, is not present in $f(R,Q)$ Palatini theories. We explicitly show that for the model (\ref{eq:f(R,Q)}) there exist bouncing solutions for which the expansion, energy density, and shear are all bounded. This model, therefore, avoids the well known problems of anisotropic universes in GR, where anisotropies grow faster than the energy density during the contraction phase leading to a singularity, which can only be avoided by means of matter sources with equation of state $w=P/\rho>1$ \cite{ekpyrotic}.

The content of the paper is organized as follows. In section \ref{sec:EOM} we summarize the field equations of Palatini $f(R,Q)$ theories with a perfect fluid, which where first derived and discussed in \cite{OSAT09}. In section \ref{sec:Exp-Shear} we obtain expressions for the expansion and shear in both $f(R)$ and $f(R,Q)$ theories. Section \ref{sec:f(R)} is devoted to the analysis of $f(R)$ theories in isotropic and anisotropic scenarios, paying special attention to the possible existence of isotropic bouncing solutions which are not of the type $\partial_Rf=0$. In section \ref{sec:f(R,Q)} we study the model (\ref{eq:f(R,Q)}) and characterize the different bouncing solutions according to the values of the Lagrangian parameters $a$ and $b$, and the equation of state $w$. 
We end with a brief discussion and conclusions. \\  

\section{Field Equations} \label{sec:EOM}

The field equations corresponding to the Lagrangian (\ref{eq:f(R,Q)}) can be derived from the action
\begin{equation}
S=\frac{1}{2\kappa^2}\int d^4x\sqrt{-g}f(R,Q)+S_m(g_{\alpha\beta},\Psi)
\end{equation}
where $R\equiv g^{\mu\nu}R_{\mu\nu}$, $Q\equiv R_{\mu\nu}R^{\mu\nu}$, $R_{\mu\nu}\equiv -\partial_{\mu}
\Gamma^{\lambda}_{\lambda\nu}+\partial_{\lambda}
\Gamma^{\lambda}_{\mu\nu}+\Gamma^{\lambda}_{\mu\nu}\Gamma^{\rho}_{\rho\lambda}-\Gamma^{\lambda}_{\nu\rho}\Gamma^{\rho}_{\mu\lambda}$,  $\Gamma^{\rho}_{\mu\lambda}$ is the independent connection, and $\Psi$ represents generically the matter fields, which are not coupled to the independent connection. Variation of the action with respect to the metric leads to
\begin{eqnarray}\label{eq:met-var}
f_R R_{\mu\nu}-\frac{f}{2}g_{\mu\nu}+2f_QR_{\mu\alpha}{R^\alpha}_\nu &=& \kappa^2 T_{\mu\nu} \ , 
\end{eqnarray}
where $f_R\equiv \partial_R f$ and $f_Q\equiv \partial_Q f$. Variation with respect to the independent connection gives 
\begin{equation}\label{eq:con-var}
\nabla_{\beta}\left[\sqrt{-g}\left(f_R g^{\mu\nu}+2f_Q R^{\mu\nu}\right)\right]=0
\end{equation}
For details on how to obtain these equations see \cite{OSAT09}. The connection equation (\ref{eq:con-var}) can be solved in general assuming the existence of an auxiliary metric $h_{\alpha\beta}$ such that (\ref{eq:con-var}) takes the form $\nabla_{\beta}\left[\sqrt{-h} h^{\mu\nu}\right]=0$. If a solution to this equation exists, then $\Gamma^{\rho}_{\mu\lambda}$ can be written as the Levi-Cività connection of the metric $h_{\mu\nu}$. When the matter sources are represented by a perfect fluid, $T_{\mu\nu}=(\rho+P)u_\mu u_\nu+P g_{\mu\nu} $, one can show that $h_{\mu\nu}$ and its inverse $h^{\mu\nu}$ are given by \cite{OSAT09}
\begin{eqnarray}
h_{\mu\nu}&=&\Omega\left( g_{\mu\nu}-\frac{\Lambda_2}{\Lambda_1-\Lambda_2} u_\mu u_\nu \right)\\
h^{\mu\nu}&=&\frac{1}{\Omega}\left( g^{\mu\nu}+\frac{\Lambda_2}{\Lambda_1} u^\mu u^\nu \right)
\end{eqnarray}
where 
\begin{eqnarray}
\Omega&=&\left[\Lambda_1(\Lambda_1-\Lambda_2)\right]^{1/2} \label{eq:def-Om}\\
\Lambda_1&=& \sqrt{2f_Q}\lambda+\frac{f_R}{2}  \label{eq:def-L1}  \\
\Lambda_2&=& \sqrt{2f_Q}\left[\lambda\pm\sqrt{\lambda^2-\kappa^2(\rho+P)}\right] \label{eq:def-L2}\\
\lambda&=&\sqrt{\kappa^2 P+\frac{f}{2}+\frac{f_R^2}{8f_Q}} \label{eq:def-lambda}\\
\end{eqnarray}
In terms of $h_{\mu\nu}$ and the above definitions, the metric field equations (\ref{eq:met-var}) take the following form
\begin{equation}\label{eq:Rmn-h}
R_{\mu\nu}(h)=\frac{1}{\Lambda_1}\left[\frac{\left(f+2\kappa^2P\right)}{2\Omega}h_{\mu\nu}+\frac{\Lambda_1\kappa^2(\rho+P)}{\Lambda_1-\Lambda_2}u_{\mu}u_{\nu}\right] \ .
\end{equation}
In this expression, the functions $f, \Lambda_1$, and $\Lambda_2$ are functions of the density $\rho$ and pressure $P$. In particular, for our quadratic model one finds that $R=\kappa^2(\rho-3P)$ and $Q=Q(\rho,P)$ is given by 
\begin{equation}\label{eq:Q}
\frac{bQ}{2R_P}=-\left(\kappa^2P+\frac{\tilde f}{2}+\frac{R_P}{8b}\tilde f_R^2\right)+\frac{R_P}{32b}\left[3\left(\frac{b R}{R_P}+\tilde f_R\right)-\sqrt{\left(\frac{b R}{R_P}+\tilde f_R\right)^2-\frac{ 4 b \kappa^2(\rho+P)}{R_P} }\right]^2 \ ,
\end{equation}
where $\tilde f=R+aR^2/R_P$, and the minus sign in front of the square root has been chosen to recover the correct limit at low curvatures. \\

In what follows, we will use (\ref{eq:Rmn-h}) to find equations governing the evolution of physical magnitudes such as the expansion, shear, matter/energy density, and so on. Note that (\ref{eq:Rmn-h}) is written in terms of the auxiliary metric  $h_{\mu\nu}$, not in terms of the physical metric $g_{\mu\nu}$. In terms of $g_{\mu\nu}$, eq. (\ref{eq:Rmn-h}) would be much less transparent and more difficult to handle. As we will see in the next section, working with (\ref{eq:Rmn-h}) will simplify many manipulations and will allow us to obtain a considerable number of analytical expressions for all the physical magnitudes of interest.

\section{Expansion and Shear} \label{sec:Exp-Shear}

In this section we derive the equations for the evolution of the expansion and shear for an arbitrary Palatini $f(R,Q)$ theory. We also particularize our results to the case of $f(R)$ theories, i.e., no dependence on $Q$. We consider a Bianchi I spacetime with physical line element of the form
\begin{equation}
ds^2=g_{\mu\nu}dx^\mu dx^\nu=-dt^2+\sum_{i=1}^3 a_i^2(t)(dx^i)^2
\end{equation} 
In terms of this line element, the non-zero components of the auxiliary metric $h_{\mu\nu}$ are the following
\begin{eqnarray}\label{eq:hmn}
h_{tt}&=& -\left(\frac{\Omega\Lambda_1}{\Lambda_1-\Lambda_2}\right)\equiv -S \\
h_{ij}&=& \Omega g_{ij}=\Omega a_i^2 \delta_{ij} 
\end{eqnarray} 
The relevant Christoffel symbols associated with $h_{\mu\nu}$ are the following
\begin{eqnarray}
\Gamma^t_{tt}&=& \frac{\dot S}{2S} \\
\Gamma^t_{ij}&=& \frac{\Omega a_i^2}{2S}\left[\frac{\dot\Omega}{\Omega}+\frac{2\dot a_i}{a_i}\right]\delta_{ij}\\
\Gamma^i_{tj}&=& \frac{\delta^i_j}{2}\left[\frac{\dot\Omega}{\Omega}+\frac{2\dot a_i}{a_i}\right]
\end{eqnarray}
The non-zero components of the corresponding Ricci tensor are
\begin{eqnarray}
R_{tt}(h)&=& -\sum_i\dot H_i-\sum_iH_i^2-\frac{3}{2}\frac{\ddot\Omega}{\Omega}+\frac{3}{4}\frac{\dot\Omega}{\Omega}\left(\frac{\dot S}{S}+\frac{\dot\Omega}{\Omega}\right)+\frac{1}{2}\left(\frac{\dot S}{S}-\frac{2\dot\Omega}{\Omega}\right)\sum_iH_i\\
R_{ij}(h)&=& \frac{\delta_{ij} a_i^2}{2}\frac{\Omega}{S}\left[2\dot H_i+\frac{\ddot\Omega}{\Omega}-\left(\frac{\dot\Omega}{\Omega}\right)^2+\frac{\dot\Omega}{\Omega}\sum_kH_k+\frac{1}{2}\frac{\dot\Omega}{\Omega}\left(\frac{3\dot\Omega}{\Omega}-\frac{\dot S}{S}\right)+\right.\nonumber \\ 
&+& \left.2H_i\left\{\sum_kH_k+\frac{1}{2}\left(\frac{3\dot\Omega}{\Omega}-\frac{\dot S}{S}\right)\right\}\right] \ ,
\end{eqnarray}
where $H_k\equiv \dot a_k/a_k$. These expressions define the Ricci tensor $R_{\mu\nu}(h)$ on the left hand side of eq. (\ref{eq:Rmn-h}). For completeness, we give an expression for the corresponding scalar curvature
\begin{equation}
R(h)= \frac{1}{S}\left[2\sum_k\dot H_k+\sum_k H_k^2+\left(\sum_k H_k\right)^2+\left(3\frac{\dot\Omega}{\Omega}-\left\{\frac{\dot S}{S}-\frac{\dot\Omega}{\Omega}\right\}\right)\sum_k H_k+3\frac{\ddot\Omega}{\Omega}-\frac{3}{2}\frac{\dot\Omega}{\Omega}\frac{\dot S}{S}\right]
\end{equation}

From the above formulas, one can readily find the corresponding ones in the isotropic, flat configuration by just replacing $H_i\to H$. For the spatially non-flat case, the $R_{tt}(h)$ component is the same as in the flat case. The $R_{ij}(h)$ component, however, picks up a new piece, $2K\gamma_{ij}$, where $\gamma_{ij}$ represents the non-flat spatial metric of $g_{ij}=a^2_i\gamma_{ij}$. The Ricci scalar then becomes $R(h)\to R^{K=0}(h)+\frac{6K}{a^2\Omega}$.\\ 

\subsection{Shear}

From the previous formulas and the field equation (\ref{eq:Rmn-h}), we find that ${R_i}^i-{R_j}^j=0$ (no summation over indices) leads to
\begin{equation}
{R_i}^i-{R_j}^j=\frac{1}{S}\left[\dot H_{ij}+H_{ij}\left\{\sum_kH_k+\frac{1}{2}\left(\frac{3\dot\Omega}{\Omega}-\frac{\dot S}{S}\right)\right\}\right]=0 \ ,
\end{equation}
where we have defined $H_{ij}\equiv H_i-H_j$. Using the matter conservation equation for a fluid with constant equation of state $P=w\rho$, 
\begin{equation}
\dot \rho=-(1+w)\rho\sum_k H_k \ ,
\end{equation}
the above equation can be readily integrated (for this reason we consider constant equations of state throughout the rest of the paper). This leads to 
\begin{equation}\label{eq:Hij}
H_{ij}=C_{ij}\frac{S^{\frac{1}{2}}\rho^{\frac{1}{(1+w)}}}{\Omega^{\frac{3}{2}}}=C_{ij}\frac{\rho^{\frac{1}{(1+w)}}}{\Lambda_1-\Lambda_2}
\end{equation}
where the constants $C_{ij}=-C_{ji}$ satisfy the relation $C_{12}+C_{23}+C_{31}=0$. It is worth noting that writing explicitly the three equations (\ref{eq:Hij}) and combining them in pairs, one can write the individual Hubble rates as follows
\begin{eqnarray}
H_1&=& \frac{\theta}{3}+\left(C_{12}-C_{31}\right)\frac{\rho^{\frac{1}{(1+w)}}}{\Lambda_1-\Lambda_2} \nonumber \\
H_2&=& \frac{\theta}{3}+\left(C_{23}-C_{12}\right)\frac{\rho^{\frac{1}{(1+w)}}}{\Lambda_1-\Lambda_2} \label{eq:Hi}\\
H_3&=& \frac{\theta}{3}+\left(C_{31}-C_{23}\right)\frac{\rho^{\frac{1}{(1+w)}}}{\Lambda_1-\Lambda_2} \nonumber
\end{eqnarray}
where $\theta$  is the expansion of a congruence of comoving observers and is defined as $\theta=\sum_i H_i$. Using these relations, the shear $\sigma^2=\sum_i\left(H_i-\frac{\theta}{3}\right)^2$ of the congruence takes the form
\begin{equation}\label{eq:shear}
\sigma^2=\frac{\rho^{\frac{2}{1+w}}}{(\Lambda_1-\Lambda_2)^2}\frac{(C_{12}^2+C_{23}^2+C_{31}^2)}{3} \ ,
\end{equation}
where we have used the relation $(C_{12}+C_{23}+C_{31})^2=0$.   

\subsection{Expansion}

We now derive an equation for the evolution of the expansion with time and a relation between expansion and shear.
From previous results, one finds that 
\begin{equation}\label{eq:Gtt}
G_{tt}(h)\equiv -\frac{1}{2}\sum_k H_k^2+\frac{1}{2}\left(\sum_k H_k\right)^2+\frac{\dot \Omega}{\Omega}\sum_k H_k+\frac{3}{4}\left(\frac{\dot \Omega}{\Omega}\right)^2
\end{equation}
In terms of the expansion and shear, this equation becomes 
\begin{equation}\label{eq:Gtt-1}
G_{tt}\equiv -\frac{\sigma^2}{2}+\frac{\theta^2}{3}\left(1+\frac{3}{2}\Delta_1\right)^2 \ , 
\end{equation}
where we have defined
\begin{equation}\label{eq:D1}
\Delta_1=-(1+w)\rho\frac{\partial_\rho\Omega}{\Omega}
\end{equation}
The right hand side of (\ref{eq:Gtt-1}) is given by $G_{\mu\nu}=\tau_{\mu\nu}-\frac{1}{2}h_{\mu\nu}h^{\alpha\beta}\tau_{\alpha\beta}$, being $\tau_{\mu\nu}$ the right hand side of (\ref{eq:Rmn-h}). A bit of algebra leads to the following relation between the expansion, shear, and the matter 
\begin{equation}\label{eq:Hubble}
\frac{\theta^2}{3}\left(1+\frac{3}{2}\Delta_1\right)^2=\frac{f+\kappa^2(\rho+3P)}{2(\Lambda_1-\Lambda_2)}+\frac{\sigma^2}{2}
\end{equation}
Note that once a particular Lagrangian is specified, an equation of state $P=w\rho$ is given, and the anisotropy constants $C_{ij}$ are chosen, the right hand side of Eqs. (\ref{eq:shear}) and (\ref{eq:Hubble}) can be parametrized in terms of $\rho$. This, in turn, allows us to parametrize the $H_i$ functions of (\ref{eq:Hi}) in terms of $\rho$ as well. This will be very useful later for our discussion of particular models. \\

In the isotropic case ($\sigma^2=0 \ , \theta=3\dot{a}/a\equiv3\Hcal$) with non-zero spatial curvature, (\ref{eq:Hubble}) takes the following form
\begin{equation}\label{eq:Hubble-iso}
\Hcal^2=\frac{1}{6(\Lambda_1-\Lambda_2)}\frac{\left[f+\kappa^2(\rho+3P)-\frac{6K\Lambda_1}{a^2}\right]}{\left[1+\frac{3}{2}\Delta_1\right]^2} 
\end{equation}

The evolution equation for the expansion can be obtained by noting that the $R_{ij}$ equations, which are of the form $R_{ij}\equiv(\Omega/2S)g_{ij}\left[\ldots\right]=(f/2+\kappa^2P)g_{ij}/\Lambda_1$, can be summed up to give 
\begin{equation}
2(\dot{\theta}+\theta^2)+\theta \left(\frac{6\dot{\Omega}}{\Omega}-\frac{\dot{S}}{S}\right)+3\left\{\frac{\ddot{\Omega}}{\Omega}+\frac{1}{2}\frac{\dot{\Omega}}{\Omega}\left(\frac{\dot{\Omega}}{\Omega}-\frac{\dot{S}}{S}\right)\right\}=\frac{3\left[f+2\kappa^2P\right]}{\Lambda_1-\Lambda_2}
\end{equation}
Using the relations $\dot{\Omega}\equiv  -(1+w)\rho\Omega_\rho \theta$, $\dot{S}\equiv  -(1+w)\rho S_\rho \theta$, and $\ddot{\Omega}=(1+w)^2\rho\theta^2[\Omega_\rho+\rho\Omega_{\rho\rho}]-(1+w)\rho \Omega_\rho \dot\theta$, the above expression turns into
\begin{equation}\label{eq:expansion}
[2+3\Delta_1]\dot\theta+\left[2+(2-3w)\Delta_1+3\Delta_2-(1+w)\rho\left(1+\frac{3}{2}\Delta_1\right)\left(\frac{{\Omega_\rho}}{\Omega}-\frac{{S_\rho}}{S}\right) \right]\theta^2=\frac{3\left[f+2\kappa^2P\right]}{\Lambda_1-\Lambda_2} \ , 
\end{equation}
where we have used the definition (\ref{eq:D1}) and have defined the quantity
\begin{equation}\label{eq:D2}
\Delta_2\equiv (1+w)^2\rho^2\frac{\Omega_{\rho\rho}}{\Omega}
\end{equation}
Note that the function $\dot\theta$ can also be plotted as a function of $\rho$. In the isotropic, non-flat case the evolution equation for the expansion ($\theta=3H$) can be obtained from (\ref{eq:expansion}) by just replacing the term $\left[f+2\kappa^2P\right]$ on the right hand side  
by $\left[f+2\kappa^2P-4K\Lambda_1/a^2\right]$.

\subsection{Limit to $f(R)$}

We now consider the limit $f_Q\to 0$, namely, the case in which the Lagrangian only depends on the Ricci scalar $R$. Doing this we will obtain the corresponding equations for shear and expansion in the $f(R)$ case without the need of extra work. 
This limit can be obtained from eqs.(\ref{eq:def-Om}) to (\ref{eq:def-lambda}) by taking $f_Q\to  0$ in those definitions. One then finds that 
\begin{eqnarray}
\Lambda_1&\to& f_R \ , \ \Lambda_2\to 0 \\
S &\to& \Omega \to f_R 
\end{eqnarray}
Equation (\ref{eq:Hij}) turns into 
\begin{equation}\label{eq:Hij-f(R)}
H_{ij}=C_{ij}\frac{\rho^{\frac{1}{(1+w)}}}{f_R}, 
\end{equation}
which leads to 
\begin{eqnarray}
H_1&=& \frac{\theta}{3}+\left(C_{12}-C_{31}\right)\frac{\rho^{\frac{1}{(1+w)}}}{f_R} \nonumber \\
H_2&=& \frac{\theta}{3}+\left(C_{23}-C_{12}\right)\frac{\rho^{\frac{1}{(1+w)}}}{f_R} \label{eq:Hi-f(R)}\\
H_3&=& \frac{\theta}{3}+\left(C_{31}-C_{23}\right)\frac{\rho^{\frac{1}{(1+w)}}}{f_R} \nonumber \ .
\end{eqnarray}
The shear in thus given by  
\begin{equation}\label{eq:shear-f(R)}
\sigma^2=\frac{\rho^{\frac{2}{1+w}}}{f_R^2}\frac{(C_{12}^2+C_{23}^2+C_{31}^2)}{3} \ ,
\end{equation}
where $C_{12}+C_{23}+C_{31}=0$. The relation between expansion and shear now becomes
\begin{equation}\label{eq:Hubble-f(R)}
\frac{\theta^2}{3}\left(1+\frac{3}{2}\tilde\Delta_1\right)^2=\frac{f+\kappa^2(\rho+3P)}{2f_R}+\frac{\sigma^2}{2}
\end{equation}
where $\tilde\Delta_1$ is given by (\ref{eq:D1}) but with $\Omega$ replaced by $f_R$. In the isotropic case with non-zero $K$ we find
\begin{equation}\label{eq:Hubble-iso-f(R)}
\Hcal^2=\frac{1}{6f_R}\frac{\left[f+\kappa^2(\rho+3P)-\frac{6K f_R}{a^2}\right]}{\left[1+\frac{3}{2}\tilde\Delta_1\right]^2} 
\end{equation}
The evolution equation for the expansion is now given by 
\begin{equation}\label{eq:expansion-f(R)}
[2+3\tilde\Delta_1]\dot\theta+\left[2+(2-3w)\tilde\Delta_1+3\tilde\Delta_2\right]\theta^2=\frac{3\left[f+2\kappa^2P-\frac{4K f_R}{a^2}\right]}{f_R} \ , 
\end{equation}
where $\tilde\Delta_2$ is defined as in (\ref{eq:D2}) but with $\Omega$ replaced by $f_R$.

\section{Isotropic and Anisotropic Bouncing $f(R)$ Cosmologies}\label{sec:f(R)}

\begin{figure}
\includegraphics[width=0.6\textwidth]{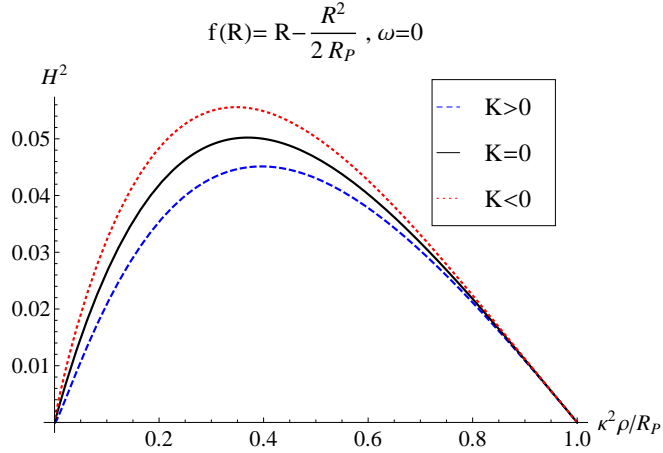}
\caption{Representation of the Hubble function in terms of $\rho$ for the model $f(R)=R-R^2/2R_P$ and $w=0$ for $K<0$, $K=0$, and $K>0$.\label{fig:fR_K}}
\end{figure}

\begin{figure}
\includegraphics[width=0.6\textwidth]{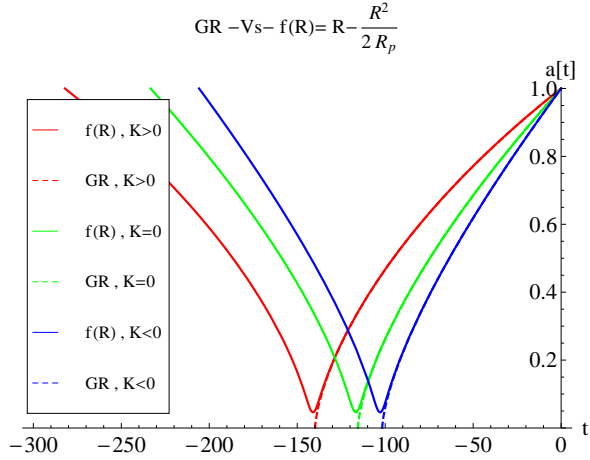}
\caption{Time evolution of the expansion factor for the model $f(R)=R-R^2/2R_P$ and $w=0$ for $K>0$, $K=0$, and $K<0$ (solid curves from left to right). From left to right, we see that the universe is initially contracting, reaches a minimum, and then bounces into an expanding phase. The dashed lines visible near the bounces represent the expanding solutions of GR, which begin with a big bang singularity and quickly tend to the non-singular solutions. \label{fig:fR_K_a}}
\end{figure}

An isotropic and homogeneous cosmological model experiences a bounce when the Hubble function $\Hcal^2$ vanishes (see Fig.\ref{fig:fR_K}), thus defining a minimum of the expansion factor (see Fig.\ref{fig:fR_K_a}). According to the formulas derived in previous sections,  isotropic bouncing $f(R)$ cosmologies occur either when the denominator of (\ref{eq:Hubble-iso-f(R)}) blows up to infinity or when the numerator $\left[f+\kappa^2(\rho+3P)-\frac{6K f_R}{a^2}\right]$ vanishes. The divergence of the denominator only depends on the form of the Lagrangian, whereas the vanishing of the numerator also depends on the value of the spatial curvature $K$. In order to characterize the anisotropic bouncing models, it is convenient to study first the isotropic case. For this reason, we will focus first on the existence of divergences in the denominator and will postpone until the end the other case. \\

\subsection{Divergences of $\tilde{\Delta}_1$ and importance of anisotropies.}

\indent As can be easily verified from the definition of $\tilde{\Delta}_1$ in (\ref{eq:D1}), the existence of divergences in the denominator of (\ref{eq:Hubble-iso-f(R)}) can only be due to the vanishing of the combination $f_R(R f_{RR}-f_R)$:
\begin{equation}
\tilde{\Delta}_1=\frac{(1+w)(1-3w)\kappa^2\rho f_{RR}}{f_R(R f_{RR}-f_R)}
\end{equation}
The Lagrangian that reproduces the dynamics of loop quantum cosmology with a massless scalar, which is well approximated by the function $f(R)=-\int dR \tanh\left(\frac{5}{103}\ln\left[\frac{R}{12R_c}\right]^2\right)$ \cite{Olmo-Singh09}, satisfies the condition $f_R=0$ at $R=12Rc$, where $R_c$ is a scale related with the Planck curvature $R_P$, thus leading to a divergence of $\tilde{\Delta}_1$ at that point. One can construct other models with simple functions such as $f(R)=R+a R^2/R_P$ or $f(R)=R+R^2/R_P(a+b\ln[R^2/R_P^2])$ which also have bounces when $f_R=0$.
 The $f_R=0$ bouncing condition seems to be quite generic and arises even when one tries to find models which satisfy the condition $R f_{RR}-f_R=0$ at some point. An illustrative example is the model $f(R)=R_P(e^{R/R_P}-1)$, which leads to $f_R=e^{R/R_P}$ and $Rf_{RR}-f_R=e^{R/R_P}(R-R_P)/R_P$, which vanishes at $R=R_P$. In this model one either finds a divergent $\Hcal^2$, due to the vanishing of the denominator of (\ref{eq:Hubble-iso-f(R)}) for $w<1/3$, or a bounce when the density approaches the limiting value $\kappa^2\rho_B=2R_P/(3w-1)$ for $w>1/3$. This bounce occurs as $R/R_P\approx \ln[1-\rho/\rho_B]\to-\infty$, which corresponds to $f_R\to 0$ and, therefore, lies in the standard class of bouncing models.\\ 

\begin{figure}
\includegraphics[width=0.65\textwidth]{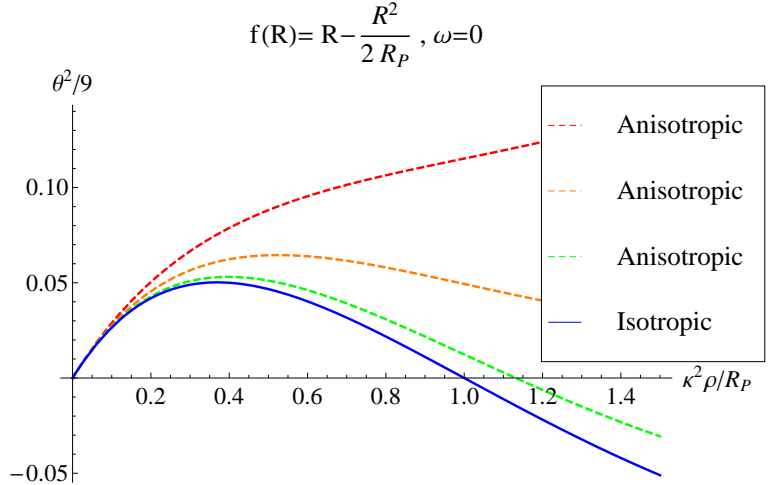}
\caption{Evolution of the expansion $\theta^2$ with the amount of anisotropy. The solid (blue) line represents the isotropic case. As the anisotropy grows, the first local maximum of $\theta^2$ disappears (dashed curves). This indicates that anisotropies can overcome the repulsive forces of the modified $f(R)$ Lagrangian. \label{fig:fR_AnisVsIso}}
\end{figure}

The importance of finding $f(R)$ models for which the bounce occurs when $f_R\neq 0$ becomes apparent when one studies anisotropic (homogeneous) scenarios. In these cases, the shear diverges as $\sim 1/f_R^2$, as is evident from (\ref{eq:shear-f(R)}). This shows that any isotropic bouncing cosmology of the $f_R=0$ type will develop divergences when anisotropies are present. And this is so regardless of how small the anisotropies are initially. It is worth noting that eventhough $\sigma^2$ diverges at $f_R=0$, the expansion and its time derivative are smooth and finite functions at that point if the density and curvature are finite. In fact, from (\ref{eq:Hubble-f(R)}) and (\ref{eq:expansion-f(R)}) we find that\footnote{Note that the case $w=1/3$ must be excluded from the analysis because in that case the theory behaves like GR with an effective cosmological constant and the manipulations that lead to (\ref{eq:bounceH}) and (\ref{eq:bounceHdot}) are not valid. }
\begin{eqnarray}\label{eq:bounceH}
\theta^2_0&=&\frac{2(C_{12}^2+C_{23}^2+C_{31}^2)}{9}\left[\frac{R_0}{(1+w)(1-3w)\kappa^2\rho_0}\right]^2\rho_0^{\frac{2}{1+w}}\\
\dot{\theta}_0&=&-\frac{(2-3w)}{3}\theta^2_0+\frac{R_0}{2(1-3w)} \label{eq:bounceHdot} \ ,
\end{eqnarray}
where the subindex denotes the point at which $f_R=0$ (where the shear diverges). It is worth noting that in GR $\dot{\theta}<0$ always, whereas in $f(R)$ the point $f_R=0$ is characterized by (\ref{eq:bounceHdot}), which may be positive, negative, or zero. If the anisotropy is sufficiently small,  which is measured by the constant $(C_{12}^2+C_{23}^2+C_{31}^2)$ in (\ref{eq:bounceH}), then $\dot{\theta}$ may be positive. This indicates that some repulsive force is trying to halt the contraction. However, if the anisotropy is too large, then it can dominate the expansion and keep $\dot{\theta}<0$ at all times (see Fig.\ref{fig:fR_AnisVsIso} and note how the first local maximum tends to disappear in the upper curves as the anisotropy grows).\\

In Fig.\ref{fig:fR_AnisVsIso} we find that there exist anisotropic solutions for which $\theta=0$ at densities beyond the point $f_R=0$, which sets the bounce of the isotropic case. One could thus be tempted to claim that for universes with low degree of anisotropy bouncing solutions really exist if we allow for slightly negative values of $f_R$, which lead to $\theta=0$. However, the shear divergences of these anisotropic models at $f_R=0$ are physically unacceptable because any detector crossing the singularity would be ripped apart by the infinite tidal forces (see \cite{Param09} for a nice discussion on divergences and singularities in cosmology). Moreover, from Eqs. (\ref{eq:Hi-f(R)}) it is easy to see that the Kretschman scalar $R_{\mu\nu\sigma\rho}R^{\mu\nu\sigma\rho}=4(\sum_i(\dot H_i+H_i^2)^2+H_1^2H_2^2+H_1^2H_3^2+H_2^2H_3^2)$ diverges at least as $\sim 1/f_R^4$, which is a clear geometrical pathology. Additionally, the vanishing of $f_R$ suggests that the field equations may not be valid for negative values of $f_R$ because the conformal transformation needed to solve for the connection becomes ill-defined at $f_R=0$, which seems to be a generic problem of anisotropic models in modified theories of gravity \cite{FF-Saa09}. Note also that the evolution of inhomogeneous perturbations in isotropic models develops divergences when $f_R$ vanishes \cite{Koivisto10}.\\

When the bounce is due to the vanishing of $R f_{RR}-f_R$ at $R=R_B$ (with $f_R\neq 0$ at that point), then the shear is finite and the expansion is given by
\begin{eqnarray}\label{eq:bounceTheta}
\theta^2_B&\sim&(R f_{RR}-f_R)^2\to 0\\
\dot{\theta}_B&=&\frac{4R_B^3f_{RRR}}{3(1+3)(1-3w)\kappa^2\rho_B}\left[\frac{f+\kappa^2(1+3w)\rho_B}{2f_R}+\frac{\sigma^2}{2}\right]_{R_B} \ . \label{eq:bounceThetadot}
\end{eqnarray}
The fact that $\theta^2_B=0$ at $R_B$ implies that the density reaches a maximum at that point (recall the conservation equation: $\dot{\rho}=-\theta(\rho+P)$). Also, since in this case the shear is finite, this family of bouncing $f(R)$ models seems to be the right family of Lagrangians to construct non-singular models. However, as we show next, there are no Lagrangians of this type able to recover GR at low curvatures. \\

\subsubsection{Non-existence of $R f_{RR}-f_R=0$ models.}

The existence of bounces in the isotropic case is due to the unbounded growth of $\left(1+\frac{3}{2}\tilde\Delta_1\right)$. One may try to build bouncing models by defining an always positive function $g(R)$ which has a divergence at $R=R_P$ such that 
\begin{equation}\label{eq:g(R)}
g(R)=2\left(1+\frac{3}{2}\tilde\Delta_1\right)=\frac{\left(f_{RR}\left[6(1+w)f-(1+3w)Rf_R\right]-f_R^2\right)}{f_R(R f_{RR}-f_R)} \ .
\end{equation}
Given the function $g(R)$, one can find the Lagrangian $f(R)$ that generates the corresponding bouncing Universe by just solving a second order differential equation. The function $g(R)$ also needs to satisfy the condition  $g(R)\approx 1$ as $R\to 0$ to force $f(R)\approx R$ in that limit. 
Simple manipulations of (\ref{eq:g(R)}) lead to
\begin{equation}\label{eq:g(R)2}
\frac{f_{RR}}{f_R^2}=\frac{[2-g(R)]}{6(1+w)f-[1+3w+g(R)]R f_R } \ .
\end{equation}
Since in GR $R>0$ if $w<1/3$ and $R<0$ if $w>1/3$, we may perform the change of variable $f(R)=\pm R_0e^{\lambda(R)}$, which leads to $f_R=\lambda_R f$, $f_{RR}=(\lambda_{RR}+\lambda_R^2)f$, and allows us to rewrite (\ref{eq:g(R)2})  as follows
\begin{equation}\label{eq:Flamb}
\frac{ \lambda_{RR} +\lambda_R^2}{\lambda_R^2}=\frac{[2-g(R)]}{6(1+w)-[1+3w+g(R)]R \lambda_R } \ . \
\end{equation}
By construction, the function $g(R)$ goes like $g(R)\approx 1$ at low curvatures, then may change in an unspecified way though always being positive at intermediate curvatures, and finally blows up to infinity at $R=R_P$, which sets the high-curvature scale. Since $g(R)$ grows unboundedly near $R_P$, we see that the denominator of (\ref{eq:Flamb}) could vanish at some point. This is in fact what one finds systematically when using a numerical trial and error scheme to find $f(R)$ bouncing models. We now show that this always occurs for any function $g(R)$ satisfying the conditions required above.
Since at low curvatures we demand $f(R)\approx R$, which implies $\lambda_R\approx 1/R$, it follows that the denominator of (\ref{eq:Flamb}) is $Den\approx 6(1+w)-(1+3w+1)1=(4+3w)$, which is positive for all reasonable matter sources ($w>-4/3$). After this initial positive value, since $g(R)>0$ will grow as $R\lambda_R$ remains positive\footnote{Note that the product $R\lambda_R$ is initially positive and can only change sign if $\lambda_R$ vanishes at some point, which would force $f_R=0$ at that point.}, unavoidably we will have $Den=0$ at some later point. Then:
\begin{itemize}
\item If $\lambda_R\neq 0$  when $Den=0$, then $g(R)$ and $\lambda_R$ are finite whereas $\lambda_{RR}\to \infty$ at that point. However, since $g(R)$ is finite, the divergence of $\lambda_{RR}$ cannot imply a cosmic bounce, since by construction that only happens when  $g(R)$ diverges. Therefore, this case does not correspond to a bounce. 
\item If we admit that $g(R)$ can indeed go to infinity, it follows that that must be the only point at which $Den=0$. This requires that the product $g(R)\lambda_R$ be finite at $R_P$, which implies that $\lambda_R\to0$ at $R_P$ to exactly compensate the divergence of $g(R)$ and give a final result which exactly cancels with the $6(1+w)$ of the denominator of (\ref{eq:Flamb}). Note that in this case the left hand side of (\ref{eq:Flamb}) diverges as $1/\lambda_R^2$ and the right hand side goes like $-g(R)/zero$.
\end{itemize}
This shows that the bouncing condition $g(R)\to \infty$ at $R_P$ can only be satisfied if $\lambda_R$ vanishes at that point, which implies that $f_R=0$ and excludes the possibility of having $Rf_{RR}-f_R=0$ as the condition for the bounce. 

\subsection{Vanishing of the numerator of $\Hcal^2$}

In the previous subsection, we concluded that the denominator of $\Hcal^2$ can only diverge if $f_R=0$. We now investigate if there exist some other mechanism able to generate isotropic bouncing models. We begin by noting that the bounce should occur when $f^B+(1+3w)\kappa^2\rho_B-6 K f_R^B/a^2_B=0$. Using the well-known relation 
\begin{equation}\label{eq:trace}
Rf_R-2f=\kappa^2T \ ,
\end{equation}
which follows from the trace of the field equations of Palatini $f(R)$ theories, we find that 
\begin{equation}
\left(R_B-\frac{12K}{a^2_B}\right)f_R^B=-3(1+w)\kappa^2\rho_B 
\end{equation}
Now, since at low curvatures $f_R\approx 1>0$ and must remain positive always (to avoid a bounce of the type $f_R=0$), at the bounce we have $(R^B-{12K}/{a^2_B})<0$ for all $w>-1$. Since at low densities $R\approx (1-3w)\kappa^2\rho$ is positive for $w<1/3$, the negative sign of $(R^B-{12K}/{a^2_B})$ implies that for $K\le 0$ the rate of growth of $R$ with $\rho$ must vanish and change sign at some point before the bounce. Using eq.(\ref{eq:trace}), we find that
\begin{equation}
\partial_\rho R=\frac{(1-3w)\kappa^2}{f_R-Rf_{RR}} \ .
\end{equation}
A change of sign in $\partial_\rho R$ implies a divergence in the denominator of this last equation, which means that $f_R\to \infty$ and/or $f_{RR}\to-\infty$. In none of those cases the theory is well defined beyond the divergence, which implies that $R$ is monotonic with $\rho$. Therefore, for $K\le0$ the only hope is a bouncing model with $w>1/3$ because for such equations of state $R<0$ always. For $K>0$ this constraint can, in principle, be avoided.\\

Let us now focus on the case $K=0$. We can parallel the strategy followed in the previous section and build $f(R)$ models starting with a function $g(R)$ which goes like $R$ at low curvatures and has a zero at $R=R_P$ such that 
\begin{equation}
\Hcal^2=\frac{g(R)}{3(1-3w)f_R\left(1+\frac{3\tilde{\Delta_1}}{2}\right)} \ .
\end{equation}
The function $g(R)$ determines a first order differential equation, $2g(R)=(1+3w)R f_R-3(1+w)f$, from which $f(R)$ can be easily obtained as
\begin{equation} 
f(R)=-\frac{2R^\gamma}{(1+3w)}\int^R dx \frac{ g(x)}{x^{1+\gamma}}
\end{equation}
where $\gamma={\frac{3(1+w)}{1+3w}}$. Though this is a convenient method for model building, a trial an error analysis does not lead to any successful model\footnote{Among many others, we considered families of models characterized by functions such as $g(R)=R(1-R^s/R_P^s)^n$ and $g(R)=R(1-(R/R_P)^s \ln R^q/R_P^q)$. }. Numerically, we find that either an $f_R=0$ bounce occurs or that the denominator of $\Hcal^2$ vanishes before the zeros of $g(R)$ can be reached, which leads to a singularity. \\

When $K\neq 0$, the above method can also be applied, though the resulting differential equation becomes highly non-linear and the solutions can only be found numerically. The results are similar to the case $K=0$. We systematically find that the models with a hope to lead to a bounce are those for which $f_R\to 0$ at some point. As a result, the spatial curvature term $-6K f_R/a^2$ is suppressed in that region and becomes negligible, giving rise to a bounce of the type $f_R=0$. Though a rigorous proof similar to that given in the case of $Rf_{RR}-f_R=0$ models is not yet available, we believe that no models of this type which recover GR at low curvatures exist. 

\subsection{Conclusions for $f(R)$ models}

Using Eqs. (\ref{eq:Hubble-f(R)}) and (\ref{eq:Hubble-iso-f(R)}), the expansion can be written as follows
\begin{equation}\label{eq:Exp_H-f(R)}
\theta^2=9H^2+\frac{3}{2}\frac{\sigma^2}{(1+\frac{3}{2}\tilde{\Delta}_1)^2} \ ,
\end{equation}
where $H$ represents the Hubble function in the $K=0$ isotropic case, and $\sigma^2$ is defined in (\ref{eq:shear-f(R)}). From this representation of the expansion, it is clear that the only way to get a true bouncing model without singularities is by satisfying the condition $R f_{RR}-f_R=0$, which would generate a finite shear, a divergent denominator in the second term of (\ref{eq:Exp_H-f(R)}), and hence a vanishing expansion. However, we have explicitly shown that such condition can never be satisfied. Moreover, even if the numerator of $H^2$ could vanish and produce a different kind of isotropic bouncing models, in the anisotropic case the expansion would not vanish and, therefore, that could not be regarded as an anisotropic bounce. For all these reasons, it follows that Palatini $f(R)$ models do not have the necessary ingredients to build a complete alternative to GR free from cosmic singularities.

\section{Nonsingular Universes in $f(R,Q)$}\label{sec:f(R,Q)}

The previous section represents a {\it no go} theorem\footnote{This is so at least for universes filled with a single perfect fluid with constant equation of state. The consideration of fluids with varying equation of state \cite{WIP} or with anisotropic stresses, see for instance \cite{Koivisto07}, could affect the dynamics adding new bouncing mechanisms, and potentially restrict the range of applicability of this conclusion.} for the existence of non-singular Palatini $f(R)$ models able to produce a complete alternative to GR in scenarios with singularities. Though the isotropic case greatly improves the situation with respect to GR, the anisotropic shear divergences kill any hopes deposited on this kind of Lagrangians. The most natural next step is to study the behavior in anisotropic scenarios of some simple generalization of the $f(R)$ family to see if the situation improves. Using the Lagrangian (\ref{eq:f(R,Q)}), we will show next that completely regular bouncing solutions exist for both isotropic and anisotropic homogeneous cosmologies. 

\subsection{Isotropic Universe}
\begin{figure}
\includegraphics[width=0.65\textwidth]{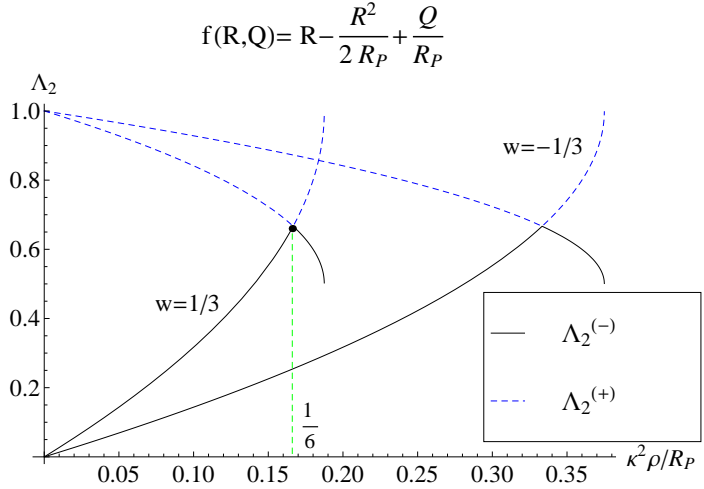}
\caption{Illustration of the need to combine the two branches of $\Lambda_2$ to obtain a continuous and differentiable curve. The branch that starts at the origin has the minus sign in front of the square root (solid line). When the square root vanishes, the function must be continued through the growing dashed branch, which corresponds to the positive sign in front of the square root. The matching point of the radiation universe ($w=1/3$) occurs at $\kappa^2\rho=R_P/6$ and appears highlighted in green (online only). \label{fig:Lambda2}}
\end{figure}

Consider Eq.(\ref{eq:Hubble-iso}) together with the definitions (\ref{eq:def-Om})-(\ref{eq:def-lambda}) particularized to the $f(R,Q)$ Lagrangian (\ref{eq:f(R,Q)}).  In this theory, we found that $R=\kappa^2(\rho-3P)$ and $Q=Q(\rho,P)$ is given by (\ref{eq:Q}). From now on we assume that the parameter $b$ of the Lagrangian is positive and has been absorbed into a redefinition of $R_P$, which is assumed positive. This restriction is necessary (though not sufficient) if one wants the scalar $Q$ to be bounded for $w>-1$. The Lagrangian then becomes $f(R,Q)=R+aR^2/R_P+Q/R_P$. When $b/R_P>0$, positivity of the square root of eq.(\ref{eq:Q}) establishes that there may exist a maximum for the combination $\rho+P$. \\ \indent The first difficulty that we find is the choice of sign in front of the square root of Eq.(\ref{eq:def-L2}). In order to recover the $f(R)$ limit and GR at low curvatures, we must take the minus sign. However, when considering particular models, which are characterized by the constant $a$ and an equation of state $w$, one realizes that the positive sign and the negative sign expressions for $\Lambda_2$ may coincide at some high curvature scale, when the argument of the square root $\sqrt{\lambda^2-\kappa^2(\rho+P)}$ vanishes. When this happens, one must make sure that the function $\Lambda_2$ at higher energies is continuous and differentiable. These two conditions force us to switch at that point from the negative to the positive sign expression (see Fig.\ref{fig:Lambda2} for an illustration of this problem), which then defines a continuous and differentiable function on the physical domain. Bearing in mind this subtlety, one can then proceed to represent the Hubble function for different choices of parameters to determine whether bouncing solutions exist or not. \\

We observe that for every value of the parameter $a$ there exist an infinite number of bouncing solutions, which depend on the particular equation of state $w$. The bouncing solutions can be divided in two large classes:
\begin{itemize}
\item {\bf Class I: $a\geq 0$.} \\ In this case, the bounce occurs at the maximum value reachable by the scalar $Q$. This happens when the argument of the square root of (\ref{eq:Q}) vanishes. In general, the density at that point satisfies
\begin{equation}\label{eq:rhoQ}
\frac{\kappa^2\rho_{Q_{max}}}{R_P}=\frac{1+5w-2a(1-3w)-\sqrt{8(1+w)(2w-a(1-3w))}}{(1+2a)^2(1-3w)^2} \ .
\end{equation}
The bounce occurs at that density for all equations of state satisfying the condition
\begin{equation}\label{eq:w-wmin}
w>w_{min}=\frac{a}{2+3a} \,
\end{equation}
which follows from the requirement of positivity of the argument of the square root of (\ref{eq:rhoQ}). 
Note that, except for $a=0$, the case $w=w_{min}$ is not contained in the set of bouncing solutions.  From (\ref{eq:w-wmin}) it follows that a radiation dominated universe, $w=1/3$, always bounces for any $a>0$. In fact, when $w=1/3$, we find that (\ref{eq:rhoQ}) must be replaced by 
\begin{equation}\label{eq:rhoQ1b3}
\kappa^2\rho_{Q_{max}^{w=1/3}}=\frac{3R_P}{16} \ .
\end{equation}
Note that this last expression is independent of the value of $a$ and, thus, holds also for the case $a\leq 0$. This was to be expected since the coefficient $a$ multiplies the quadratic term $R^2/R_P$, which is zero in a radiation dominated universe. Remarkably, this implies that all radiation dominated universes in the family of Lagrangians considered here always lead to a big bounce. This clearly demonstrates that $f(R,Q)$ theories posses interesting dynamical properties that cannot be reproduced by any $f(R)$ Palatini Lagrangian \cite{OAST09b}. The modified dynamics in the $f(R)$ case is generated by new terms that depend on the trace $T=-(1-3w)\rho$, which do not produce any effect in a radiation scenario.\\
  
\item {\bf Class II: $a\leq 0$.}\\ This case is more involved and must be divided in several intervals. In general, the bounce occurs at a density given by the following expression
\begin{equation}\label{eq:rho_B}
\frac{\kappa^2\rho_{B}}{R_P}=\left\{\begin{tabular}{lr} $\frac{1+6w-2a(1-3w)-3\sqrt{w(2+3w)-a(1+w)(1-3w)}}{(1+a)(1+4a)(1-3w)^2}$ & \mbox{ if}  $w \leq w_{0}$\\ 
                                                $\frac{\kappa^2\rho_{Q_{max}}}{R_P}$ & \mbox{ if } $w\geq w_{0}$ \end{tabular}\right.
\end{equation}
where $w_0$ represents the value of $w$ at which the two branches of $\kappa^2\rho_B/R_P$ coincide. The generic expression for $w_0$ as a function of $a$ is very complicated, though its computation for a given $a$ is straightforward. Note that the curve defined by Eq.(\ref{eq:rho_B}) is smooth and differentiable with respect to $w$ even at $w_0$. It is important to note that $w_0$ is always negative. This means that the bouncing solutions that occur at $\rho_{Q_{max}}$ can be extended to negative values of $w$ until the value $w_0$. As of that point, the range of bouncing solutions is extended to even more  negative values of $w$ through the new branch $w\leq w_0$ of Eq.(\ref{eq:rho_B}). What happens before and after $w_0$ to make that particular equation of state so relevant? The answer is as follows. For $w\geq w_0$, the bounce occurs at a density for which $Q$ is maximum (when the square root of (\ref{eq:Q}) vanishes). For $w\leq w_0$, the bounce occurs at a density for which the function $\Lambda_1-\Lambda_2$ vanishes. At $w=w_0$ we find that $Q$ reaches its maximum at the same density as $\Lambda_1-\Lambda_2$ vanishes. \\

How far into the negative axis can $w$ be extended beyond the matching point $w_0$? The answer depends on the value of $a$. We split the $a<0$ axis in five elements:
\begin{itemize}
\item {\bf Case IIa: $-1/4< a\leq 0$}\\
The values of $w$ in this interval are restricted by the argument of the square root of (\ref{eq:rho_B}) for $w\leq w_0$. We thus find that 
\begin{equation}
-\frac{1}{3}+\frac{1}{3}\sqrt{\frac{1+4a}{1+a}}<w<\infty
\end{equation}
We see that when $a=0$ we find agreement with the discussion of Case I. As $a$ approaches the limiting value $-1/4$, the bouncing solutions extend up to $w\to -1/3$. However, since the branch $w\leq w_0$ of (\ref{eq:rho_B}) is singular at $a=-1/4$, that particular model must be studied separately.

\item {\bf Case IIb: $a=-1/4$} \\
In this case, the density at the bounce is given by the following expression
\begin{equation}\label{eq:rho_B_a1/4}
\frac{\kappa^2\rho_{B}}{R_P}=\left\{\begin{tabular}{lr} $\frac{1}{3(1+3w)}$ & \mbox{ if}  $w \leq -\frac{1}{9}$\\ 
                                                $\frac{\kappa^2\rho_{Q_{max}}}{R_P}$ & \mbox{ if } $w\geq -\frac{1}{9}$ \end{tabular}\right.
\end{equation}
which is always finite except for the limiting value $w=-1/3$. Thus, bouncing solutions exist for any $w$ within the interval $-1/3\leq w<\infty$. 

\item {\bf Case IIc: $-1/3\leq a\leq-1/4$}\\
Though in this interval the argument of the square root in (\ref{eq:rho_B}) is always positive, we observe numerically that the bouncing solutions cannot be extended beyond the value $w<-1$, where $\rho_B$ reaches a maximum. Therefore, in this interval we find that the bouncing solutions occur if $-1<w<\infty$, where $w=-1$ is excluded. 

\item {\bf Case IId: $-1\leq a\leq-1/3$}\\
Here we also find that the negative values of $w$ cannot be extended beyond $w<-1$. Surprisingly, we also find restrictions for $w>1$ which are due to the existence of zeros in the denominator of $H^2$. Due to the algebraic complexity of the functions involved, it is not straightforward to find a clean way to characterize the origin of those zeros. However, numerically we find that they arise when $w\approx(\alpha+\beta a)/(1+3a)^2$, where $\alpha=1.1335$ and $\beta=-3.3608$ (this fit is very good near $a\approx -1/3$ and slightly worsens as we approach $a=-1$). Summarizing, the bouncing solutions are restricted to the interval $-1<w< (\alpha+\beta a)/(1+3a)^2>1$. This expression agrees in the limiting value $a=-1/3$ with the expected values $-1<w<\infty$ of the case IIc. The case $a=-1$ must be treated separately, though it does not present any undesired feature. \\
Note that in this interval one finds the case $a=-1/2$, which is singular according to (\ref{eq:rhoQ}) and must be treated separately. We find that Eq. (\ref{eq:rhoQ}) must be replaced by $\kappa^2\rho_{Q_{max}}=1/(4+4w)$. Other than that, this case satisfies the same rules as the other models in this interval. 

\item {\bf Case IIe: $a\leq-1$} \\
Similarly as the family $a\geq 0$, this set of models also allows for a simple characterization of the bouncing solutions, which correspond to the interval $-1<w<a/(2+3a)$. In the limiting case $a=-1$ we obtain the condition $-1<w<1$ (compare this with the numerical fit above, which gives $-1<w<1.12$). In that case, the density at the bounce is given by 
\begin{equation}\label{eq:rho_B_a-1}
\frac{\kappa^2\rho_{B}}{R_P}=\left\{\begin{tabular}{lr} $\frac{1}{6}$ & \mbox{ if}  $w \leq -\frac{1}{3}$\\ 
                                                $\frac{\kappa^2\rho_{Q_{max}}}{R_P}$ & \mbox{ if } $w\geq -\frac{1}{3}$ \end{tabular}\right.
\end{equation}
For $a<-1$, the equations of state that generate bouncing solutions get reduced from the right and approach  $-1<w\leq1/3$ as $a\to-\infty$, with the case $w=1/3$ always included. 

\end{itemize}

\end{itemize}

\subsection{Anisotropic Universe}

Using Eqs. (\ref{eq:Hubble}) and (\ref{eq:Hubble-iso}), the expansion can be written as follows
\begin{equation}\label{eq:Exp_H}
\theta^2=9H^2+\frac{3}{2}\frac{\sigma^2}{(1+\frac{3}{2}\Delta_1)^2} \ ,
\end{equation}
where $H$ represents the Hubble function in the $K=0$ isotropic case. To better understand the behavior of $\theta^2$, let us consider when and why $H^2$ vanishes. Using the results of the previous subsection, we know that $H^2$ vanishes either when the density reaches the value $\rho_{Q_{max}}$ or when the function $\Lambda_1-\Lambda_2$ vanishes. These two conditions imply a divergence in the quantity $(1+\frac{3}{2}\Delta_1)^2$, which appears in the denominator of $H^2$ and, therefore, force the vanishing of $H^2$ (isotropic bounce). Technically, these two types of divergences can be easily characterized. From the definition of $\Delta_1$ in (\ref{eq:D1}), one can see that $\Delta_1\sim \partial_\rho\Omega/\Omega$. Since $\Omega\equiv \sqrt{\Lambda_1(\Lambda_1-\Lambda_2)}$, it is clear that $\Delta_1$ diverges when $\Lambda_1-\Lambda_2=0$. The divergence due to reaching $\rho_{Q_{max}}$ is a bit more elaborate. One must note that $\partial_\rho\Omega \sim \partial_\rho\Lambda_1\sim \partial_\rho\Lambda_2$ and that $\partial_\rho\Lambda_{1,2}$ contain terms that are finite plus a term of the form $\partial_\rho\lambda$, with $\lambda$ given by (\ref{eq:def-lambda}). In this $\lambda$ there is a $Q-$term hidden in the function $f(R,Q)$, which implies that $\partial_\rho\lambda\sim \partial_\rho Q/R_P$ plus other finite terms. From the definition of $Q$ it follows that $\partial_\rho Q$ has finite contributions plus the term $\partial_\rho \Phi /\sqrt{\Phi}$, where $\Phi\equiv(1+(1+2a)R/R_P)^2-4\kappa^2(\rho+P)/R_P$, which diverges when $\Phi$ vanishes. This divergence of $\partial_\rho Q$ indicates that $Q$ cannot be extended beyond the maximum value $Q_{max}$. \\
\indent Now, since the shear goes like $\sigma^2\sim 1/(\Lambda_1-\Lambda_2)^2$ [see Eq.(\ref{eq:shear})], we see that the condition $\Lambda_1-\Lambda_2=0$ implies a divergence on $\sigma^2$ (though $\theta^2$ remains finite). This is exactly the same type of divergence that we already found in the $f(R)$ models. In fact, the decomposition (\ref{eq:Exp_H}) is also valid in the $f(R)$ case, where $\Lambda_2\to 0$ and $\Lambda_1\to f_R$ (see Eq.(\ref{eq:Exp_H-f(R)})). Since in those models the bounce can only occur when $f_R=0$, which is equivalent to the condition $\Lambda_1-\Lambda_2=0$, there is no way to achieve a completely regular bounce using an $f(R)$ theory. On the contrary, since the quadratic $f(R,Q)$ model (\ref{eq:f(R,Q)}) allows for a second mechanism for the bounce, which takes place at $\rho_{Q_{max}}$, there is a natural way out of the problem with the shear. \\

When the density reaches the value $\rho_{Q_{max}}$, we found in the previous subsection that the combination $\Lambda_1-\Lambda_2$ is always greater than zero  except for the particular equation of state $w=w_0$ (recall that $w_0$ was defined as the matching condition in eq.(\ref{eq:rho_B}), and represents the case in which $\rho_{Q_{max}}$ is reached at the same time as $\Lambda_1-\Lambda_2=0$). Therefore, for any $w> w_0$ the shear will always be finite at $\rho_{Q_{max}}$. Moreover, since at that point the denominator $(1+\frac{3}{2}\Delta_1)^2$ blows up to infinity, it follows that the expansion vanishes there, which sets a true maximum for $\rho$ like in the isotropic case. \\
At this point one may wonder about the consequences of the divergence of $\partial_\rho Q$ at $\rho_{Q_{max}}$ for the consistency of the theory. This question is pertinent because the connection that defines the Riemann tensor involves derivatives of $\Omega$ and hence of $Q$. In this sense, it should be noted that because of the spatial homogeneity only time derivatives of such quantities need be considered. We are thus interested in objects such as $\partial_t Q$ and higher time derivatives. One can check by direct computation that $\partial_t Q=(\partial_\rho Q)\dot{\rho}$ yields a finite result because the divergence of $\partial_\rho Q$ is exactly compensated by the vanishing of $\dot{\rho}$, which is due to the vanishing of the expansion at the bounce. Explicit computation of higher derivatives of $Q$ and other relevant objects (such as $\Omega$, $S$, \ldots \ needed to compute the components of the Ricci tensor) shows that all them are well behaved at the point of the bounce\footnote{This same reasoning can be used to confirm the pathological character of the other type of bounce, the one characterized by the condition $\Lambda_1-\Lambda_2=0$.}. This guarantees that the bounce is a completely regular point that does not spoil the well-posedness of the time evolution nor the disformal transformation needed to relate the physical and the auxiliary metrics $g_{\mu\nu}$ and $h_{\mu\nu}$, respectively.  \\
	
	Summarizing, we conclude that the Lagrangian (\ref{eq:f(R,Q)}) leads to completely regular bouncing solutions in the anisotropic case for $w>\frac{a}{2+3a}$ if $a\geq 0$, for $w_0<w<\infty$ if $-1/3\leq a\leq 0$, for $w_0<w<(\alpha+\beta a)/(1+3a)^2$ if $-1\leq a\leq-1/3$, and for $-1/3<w<a/(2+3a)$ if $a\leq-1$, where $w_0<0$ is defined using (\ref{eq:rho_B}) and its corresponding subcases. These results imply that for $a<0$ the interval $0\leq w\leq 1/3$ is always included in the family of bouncing solutions, which contain the dust and radiation cases. For $a\geq 0$, the radiation case is always non-singular too. 

\subsection{An example: radiation universe.}

\begin{figure}
\includegraphics[width=0.65\textwidth]{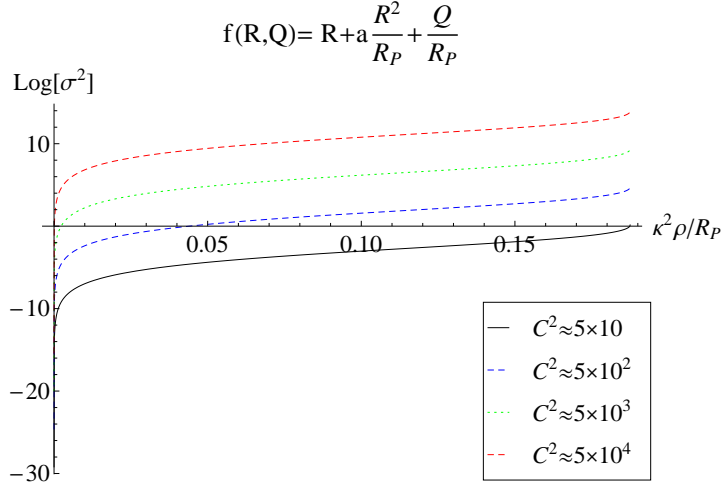}
\caption{Logarithmic representation of the shear as a function of $\kappa^2\rho/R_P$ in radiation universes with different value of the anisotropy, which is controlled by the combination $C^2=C_{12}^2+C_{23}^2+C_{31}^2$. In this representation, the difference between the curves is just a constant shift of magnitude $\log C^2$.  \label{fig:ShearRad}}
\end{figure}

\begin{figure}
\includegraphics[width=0.65\textwidth]{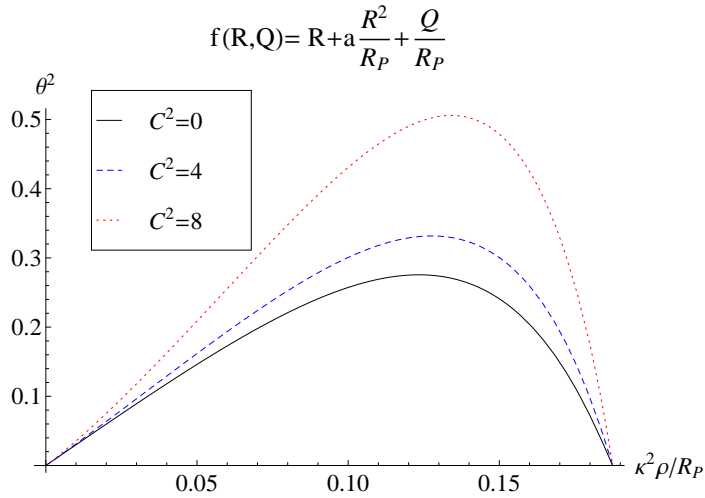}
\caption{Evolution of the expansion as a function of $\kappa^2\rho/R_P$ in radiation universes with low anisotropy, which is controlled by the combination $C^2=C_{12}^2+C_{23}^2+C_{31}^2$. The case with $C^2=0$ corresponds to the isotropic flat case, $\theta^2=9H^2$.  \label{fig:ExpanRad}}
\end{figure}

\begin{figure}
\includegraphics[width=0.65\textwidth]{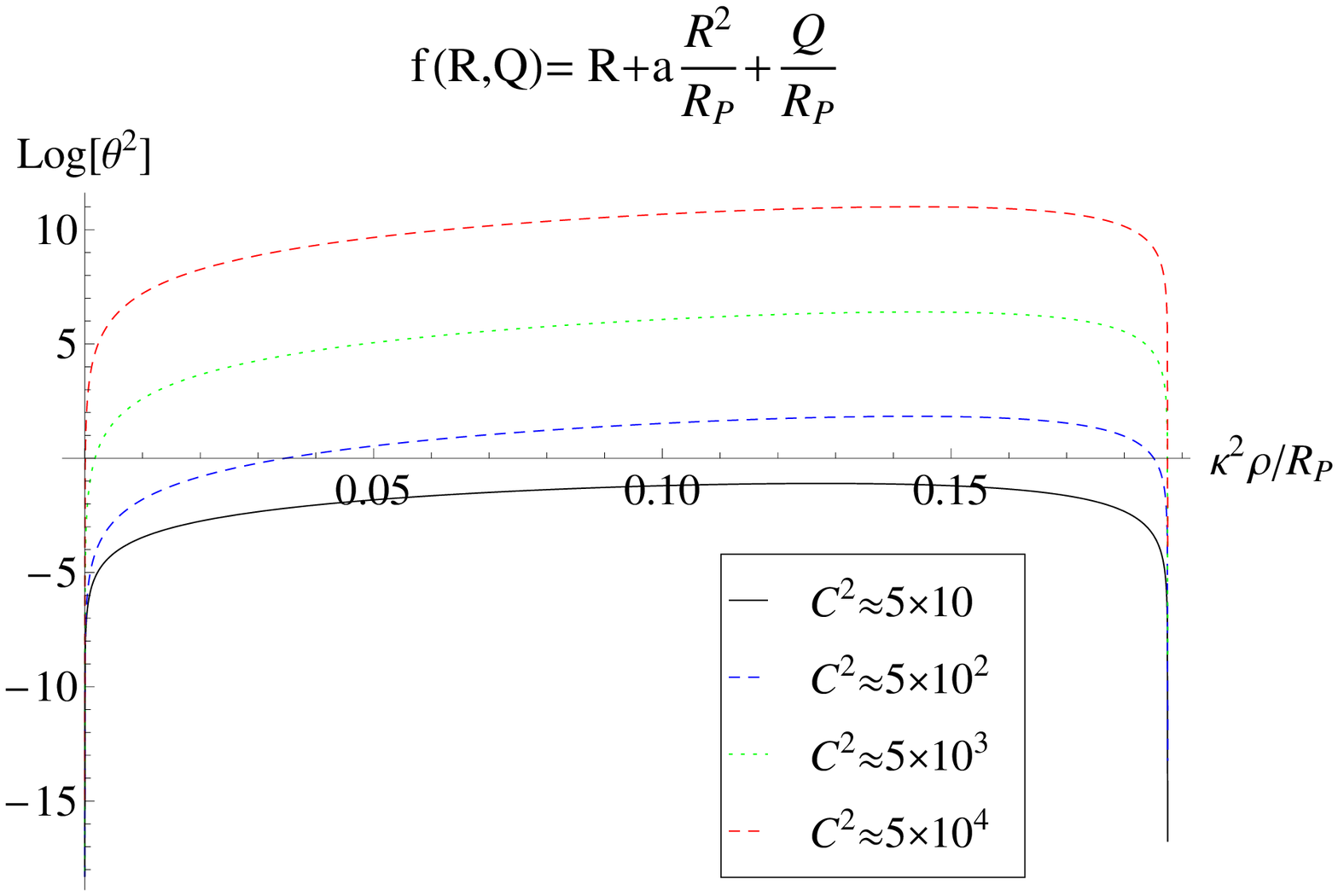}
\caption{Logarithmic representation of the expansion as a function of $\kappa^2\rho/R_P$ in radiation universes with high anisotropy, which is controlled by the combination $C^2=C_{12}^2+C_{23}^2+C_{31}^2$. \label{fig:LExpanRad}}
\end{figure}

As an illustrative example, we consider here the particular case of a universe filled with radiation. Besides its obvious physical interest, this case leads to a number of algebraic simplifications that make more transparent the form of some basic definitions
\begin{eqnarray}
Q&=& \frac{3R_P^2}{8}\left[1-\frac{8\kappa^2\rho}{3R_P}-\sqrt{1-\frac{16\kappa^2\rho}{3R_P}}\right] \label{eq:Q-rad}\\
\lambda &=& \frac{3}{4}\left(1-\frac{1}{3}\sqrt{1-\frac{16\kappa^2\rho}{3R_P}}\right)\sqrt{\frac{R_P}{2}} \\
\Lambda_1 &=& \frac{1}{2}+\frac{3}{4}\left(1-\frac{1}{3}\sqrt{1-\frac{16\kappa^2\rho}{3R_P}}\right) 
\end{eqnarray}
It is easy to see that the coincidence of the two branches of $\Lambda_2$ occurs at $\kappa^2\rho=R_P/6$. Therefore, the physical $\Lambda_2$ must be defined as follows (see Fig.\ref{fig:Lambda2})
\begin{equation}\label{eq:L2rad}
\Lambda_2=\left\{\begin{tabular}{lr} $\frac{1}{\sqrt{8}}\left[\sqrt{5-\frac{8\kappa^2\rho}{3R_P}-3\sqrt{1-\frac{16\kappa^2\rho}{3R_P}}}-\sqrt{5-\frac{24\kappa^2\rho}{R_P}-3\sqrt{1-\frac{16\kappa^2\rho}{3R_P}}} \right] $ & \mbox{ if}  $\kappa^2\rho \leq \frac{R_P}{6}$\\                                 \\     $\frac{1}{\sqrt{8}}\left[\sqrt{5-\frac{8\kappa^2\rho}{3R_P}-3\sqrt{1-\frac{16\kappa^2\rho}{3R_P}}}+\sqrt{5-\frac{24\kappa^2\rho}{R_P}-3\sqrt{1-\frac{16\kappa^2\rho}{3R_P}}} \right] $ & \mbox{ if}  $\kappa^2\rho \geq \frac{R_P}{6}$ \end{tabular}\right.
\end{equation}
This definition by parts unavoidably obscures the representation of other derived quantities. Nonetheless, it is necessary to obtain continuous and differentiable expressions for the physical magnitudes of interest such as the expansion and shear (plotted in Figs. \ref{fig:ShearRad}, \ref{fig:ExpanRad}, and \ref{fig:LExpanRad}). It is easy to see that at low densities (\ref{eq:Q-rad}) leads to $Q\approx 4(\kappa^2\rho)^2/3+32(\kappa^2\rho)^3/9R_P+320(\kappa^2\rho)^4/27R_P^2+\ldots$, which recovers the expected result for GR, namely, $Q=3P^3+\rho^2$. From this formula we also see that the maximum value of $Q$ occurs at $\kappa^2\rho_{max}=3R_P/16$ and leads to $Q_{max}=3R_P^2/16$. At this point the shear also takes its maximum allowed value, namely, $\sigma^2_{max}=\sqrt{3/16}R_P^{3/2}(C_{12}^2+C_{23}^2+C_{31}^2)$, which is always finite. At $\rho_{max}$ the expansion vanishes producing a cosmic bounce regardless of the amount of anisotropy.

\section{Discussion and conclusions}   

In this work we have shown that simple modifications of GR with high curvature corrections in Palatini formalism successfully avoid the big bang singularity in isotropic and anisotropic (Bianchi-I) homogeneous cosmologies giving rise to bouncing solutions. And this type of solutions seem to be the rule rather than the exception. The $f(R,Q)$ model (\ref{eq:f(R,Q)}) in Palatini formalism is just an example. This type of models is motivated by the fact that the effective dynamics of loop quantum cosmology \cite{lqc} is described by second-order equations and by the need to go beyond the dynamics of Palatini $f(R)$ theories, which cannot avoid the development of shear singularities in anisotropic scenarios (as has been shown in section IV).\\ 

In the model (\ref{eq:f(R,Q)}), regular sources of matter and radiation can remove the singularities thanks to the unconventional interplay between the matter and the geometry at very high energies. Due to the form of the gravity Lagrangian (\ref{eq:f(R,Q)}), at low energies the theory recovers almost exactly the dynamics of GR because the connection coincides with the Levi-Civita connection of the metric up to completely negligible corrections of order $\sim \kappa^2\rho/R_P$. At high energies, however, the departure is significant and that results in modified dynamics that resolves the singularity. The assumption that metric and connection are regarded as independent fields (Palatini variational principle) is at the root of this phenomenon, which could provide new insights on the properties of the quantum geometry and its interaction with matter. Because of this independence between metric and connection, the dynamics of our model turns out to be governed by second-order equations. As a result,  the avoidance of the big bang singularity is not due to the existence of multiple new solutions of the field equations suitably tuned to get the desired result. Rather, the physically disconnected contracting and expanding solutions found in GR, which end or start in singularities, are suitably {\it deformed} due to the non-linear dependence of the expansion on the matter/radiation density and produce a single regular branch (this non-linear density dependence is also manifest in loop quantum cosmology \cite{lqc} and has recently been identified in \cite{Bozza-Bruni09} as a possible solution to the anisotropy problem). At low energies, the standard solutions of GR are smoothly recovered, and such solutions uniquely determine the high energy behavior. This should be contrasted with the same $f(R,Q)$ Lagrangian formulated in the metric formalism, where due to the existence of additional degrees of freedom multiple new solutions arise and one must use an {\it ad hoc} procedure to single out those which recover a FRW expansion at late times. \\

The fact that for all negative values of the parameter $a$ one finds isotropic and anisotropic bouncing solutions in universes filled with dust 
indicates that with the Palatini modified dynamics the mere presence of matter is enough to significantly alter the geometry to avoid the singularity. 
Unlike in pure GR, there is no need for exotic sources of matter/energy with unusual interactions or unnatural equations of state. Regular matter is able by itself to generate repulsive gravity when a certain high energy scale is reached. And this occurs in a non-perturbative way. In fact, in the case of a radiation universe, for instance, the perturbative expansion of $Q$ (see below equation (\ref{eq:L2rad})) does not suggest the presence of any significant new effect as the scale $R_P$ is approached. However, a glance at the exact expression (\ref{eq:Q-rad}) shows that there exists a maximum value for $\rho$, which is set by the positivity of the argument of the square root. Such limiting value is only apparent when the infinite series expansion of $Q$ is explicitly considered. It is interesting to note that this type of non-perturbative effect arises in our theory without the need for introducing new dynamical degrees of freedom. In other approaches to non-singular cosmologies, the non-perturbative effects are introduced at the cost of adding an infinite number of derivative terms in the action (see \cite{Biswas:2010zk} for a recent example).\\
 Additionally, since in radiation dominated universes ($w=1/3$) the scalar curvature vanishes, $R=0$, the mechanism responsible for the bounce in these models is directly connected with the $Q=R_{\mu\nu}R^{\mu\nu}$ term of the Lagrangian. In fact, all Lagrangians of the form $f(R,Q)=\tilde{f}(R)+Q/R_P$, will lead to the same cosmic dynamics\footnote{In radiation scenarios, all $\tilde{f}(R)$ functions which satisfy $\partial_R\tilde{f}(0)=1$ will lead to the same dynamics up to an effective cosmological constant, which we assume very small and negligible during the very early universe.} if $w=1/3$, as is easy to see from the definitions (\ref{eq:def-Om})-(\ref{eq:def-lambda}) and (\ref{eq:Q}). This is a clear indication of the robustness of the models $f(R,Q)=\tilde{f}(R)+Q/R_P$ against cosmic singularities. Note, in addition, that the anisotropic bounce always occurs when the maximum value of $Q$ is reached, which emphasizes the crucial role of this term in the dynamics.\\
On the other hand, the fact that this class of Palatini $f(R,Q)$ actions can keep anisotropies under control for a very wide range of equations of state (including radiation and dust) without the need for introducing exotic sources (as in ekpyrotic models, which require $w>1$), turns these theories into a particularly  interesting alternative to non-singular inflationary models. \\

To conclude, our investigation of anisotropies in Palatini $f(R)$ and $f(R,Q)$ models has been very fruitful. On the one hand we have been able to identify serious limitations of the $f(R)$ models in anisotropic scenarios, namely, the existence of generic shear divergences, which makes these models unsuitable for the construction of fully viable alternatives to GR. On the other hand, we have shown that the model (\ref{eq:f(R,Q)}) is a good candidate to reach the goal of building a singularity free theory of gravity without adding new dynamical degrees of freedom. Whether this particular model can successfully remove singularities in more general spacetimes is a matter that will be studied in future works.\\

\noindent { \bf Acknowledgements.} This
work has been partially supported by the Spanish grants FIS2008-06078-C03-02, FIS2008-060078-C03-03, and  the  Consolider-Ingenio 2010 Programme CPAN (CSD2007-00042). G.O. thanks MICINN for a JdC contract and the ``Jos\'e Castillejo'' program for funding a stay at the University of Wisconsin-Milwaukee, where part of this work was carried out. The authors are grateful to H. Sanchis-Alepuz for continuous and stimulating discussions on several aspects of this work, and to F. Barbero, T. Koivisto, and G. Mena-Marugán for useful comments, suggestions, and criticisms.

\end{document}